\documentclass[pra,showpacs,citeautoscript,twocolumn]{revtex4}
\usepackage{graphicx}
\usepackage{amsmath}

\begin{document}
\title{Influence of Low Frequency Noise on
Macroscopic Quantum Tunneling in Superconducting Circuits}

\begin{abstract}
The influence of low to moderate frequency environments on
Macroscopic Quantum Tunneling (MQT) in superconducting
circuits is studied within the Im$F$ approach to evaluate
tunneling rates. Particular attention is paid to two model
environments, namely, a pure sluggish bath and a sluggish
bath with additional $1/f$ noise. General findings are
applied to Zener flip tunneling, a MQT phenomenon
recently predicted and observed in a superconducting
circuit implementing a quantum bit.
\end{abstract}


\author{Mathias Duckheim$^{1,\dagger}$}
\email{mduck@tfp1.uni-freiburg.de}
\author{Joachim Ankerhold$^{1,2}$}
\affiliation{$^1$Physikalisches Institut,
Albert-Ludwigs-Universit{\"a}t Freiburg,
Hermann-Herder-Stra{\ss}e 3, D-79104 Freiburg, Germany,
\break
 $^2$Quantronics group, Service de Physique de l'Etat
Condens\'{e}, DSM/DRECAM, CEA Saclay, 91191\
Gif-sur-Yvette, France}
\date{\today}
\maketitle

\section{Introduction}

The decay of a metastable system through macroscopic
quantum tunneling (MQT) has been studied in view of
conceptual questions in quantum theory
 already in the 1980s \cite{caldeira_leggett,devoret}. Recently, it
has regained new interest e.g.\ in the context of quantum
information processing in solid state based systems
\cite{martinis0,denis} or as an effective detection
mechanism in shot noise measurements \cite{nazarov}. In
corresponding electrical circuits, current-biased
Josephson junctions are used as building blocks where the
phase difference across a junction is the only relevant
degree of freedom. The switching from the zero voltage
state can then be visualized as the dynamics of a
fictitious particle moving in a cubic potential such that
at low temperatures the escape is dominated by quantum
tunneling through the potential barrier. The crucial
impact of the electromagnetic environment on this process
has already been elucidated twenty years ago and then led
to the development of the "standard" theory of quantum
dissipative systems \cite{weiss}.

To date, the most powerful approach to calculate tunneling
rates in dissipative systems is the so-called Im$F$-method
which relates the escape rate to the imaginary part of the
free energy of an unstable system
\cite{langer,caldeira_leggett}. Basically, it can be seen
as the generalization of the method to extract the lifetime
of an unstable state from an imaginary part of its
resonance energy \cite{waxman_leggett}. The theoretical
predictions of the Im$F$ theory including dissipation
\cite{olschowski} have been thoroughly tested in the 1980s
in Josephson junction circuits \cite{devoret}. In these
studies the dominant effect of friction was due to
dynamical environmental modes with ohmic spectral density.
These modes generate at very low temperatures $T$  a rate
enhancement proportional to $T^2$ as compared to the zero
temperature limit \cite{thermalenhancement,martinis}.

In recent experiments with superconducting circuits for the
implementation of quantum bits, the role of low frequency
modes in the environmental spectrum has turned out to be
substantial
\cite{paladino,shnirman,paladino2,clarke,averin}. Bath
modes that are inert on the typical experimental time
scales or show only slow to moderate dynamics basically
determine the dephasing time for oscillations of coherent
superpositions of qubit states. In many cases, the
corresponding noise spectrum displays a $1/f$ dependence
over a broad frequency range attributed e.g.\ to low
frequency bistable charge fluctuators or electromagnetic
fluctuations in control lines. In fact, $1/f$ noise has
been known for years to be always present in mesoscopic
devices \cite{dutta,zorinahlers}. While detailed
theoretical studies on its influence on the decoherence
process in qubit devices have been given, apart from some
qualitative estimates, much less is known about its impact
on MQT.

In this paper, we analyse MQT processes  subject to noise
from very low up to moderate frequencies. One may expect
that
 quantum fluctuations in the bath are  thus
suppressed. For two model environments, namely, a pure
sluggish bath and an environment with $1/f$
characteristic, we show in detail to what extent this is
actually true and how such a situation is incorporated into
the Im$F$ theory. These findings are further applied to a
particular kind of MQT phenomenon which has  been predicted
\cite{ankerhold} and observed \cite{zener2} recently in
the so-called quantronium circuit \cite{denis}. This
system, a superconducting circuit implementing a two-level
system (qubit) with a readout by MQT, realizes an extension
of the standard MQT scenario: It describes a fictitious
particle with a spin-$\frac{1}{2}$ as an internal degree
of freedom such that the Zeeman splitting of its levels is
position dependent; in certain ranges of parameter space
the Zeeman levels cross under the barrier leading to spin
flips while tunneling. The MQT rate is then substantially
enhanced since effectively  the particle has to penetrate a
smaller barrier.

The article is organized as follows. In Sec.~\ref{imf} we
give a brief account of  how tunneling rates in dissipative
quantum systems are calculated within the Im$F$ approach.
Then, for the generic case of a cubic potential the two
models of low frequency environments are analysed, a
sluggish bath model in Sec.~\ref{adiabatic}, and a model
that incorporates additional dynamical modes with a $1/f$
characteristic in Sec.~\ref{dynamic}. The MQT readout in
the quantronium device is addressed in
Sec.~\ref{quantronium}, particularly the Zener flip MQT.
At the end some conclusions are given.

\section{Free energy method for tunneling rates}\label{imf}

The stochastic motion of a one-dimensional degree of
freedom $q$ with mass $M$ in a potential  field $V(q)$ is
determined by a classical  Langevin equation
\begin{equation}
M\ddot{q}+M \int_0^t ds\, \gamma(t-s)
\dot{q}(s)+\frac{\partial V}{\partial q}=\xi(t)\, .
\label{eq1-1}
\end{equation}
The damping kernel $\gamma(t)$ and the Gaussian fluctuating
force with $\langle \xi(t)\rangle=0$ are related by the
fluctuation-dissipation theorem $\langle
\xi(t)\xi(s)\rangle=M k_{\rm B} T \gamma(t-s)$. The
situation where a particle is initially confined in a
metastable well and, as time elapses, leaves it due to
thermal activation has been studied in a variety of
systems \cite{hanggirep,weiss}. In the low temperature
range below a certain crossover temperature $T_0$, however,
quantum tunneling is known to be the dominant escape
process. Particularly, for an undamped system one has
$T_0=\hbar\omega_0/(2\pi k_{\rm B})$ where $\omega_0$
denotes the frequency for oscillations around the minimum
of the well (plasma frequency).

To describe quantum  barrier penetration in presence of a
dissipative environment, one starts from a
system+reservoir model $H=H_0+H_{\rm B}+H_{\rm I}$, where
\begin{eqnarray}
H_0 &=&\frac{p^2}{2M}+V(q)\nonumber\\
H_{\rm B} &=& \sum_i \frac{p_i^2}{2 m_i}+\frac{1}{2}
m_i\omega_i^2 x_i^2\nonumber\\
H_{\rm I}&=& -q \sum_i c_i x_i \label{eq1-2}
\end{eqnarray}
This model of a bath consisting of harmonic oscillators
captures linear dissipation and equivalently mimics an
environment with Gaussian statistics. In the limit of a
quasi-continuum of oscillators the effective influence of
the heat bath is determined by the temperature $T$ and the
spectral bath density $J(\omega)$. Classically, for this
model one regains the Langevin equation (\ref{eq1-1})
where the Laplace transform of the damping kernel
\begin{equation}
\hat{\gamma}(z)=\int_0^\infty dt \gamma(t) {\rm e}^{-z
t}\label{eq1-3}
\end{equation}
is related to the spectral density via
\begin{equation}
\hat{\gamma}(z)=\frac{1}{M}\int_0^\infty
\frac{d\omega}{\pi} \frac{J(\omega)}{\omega}\, \frac{2
z}{\omega^2+z^2}\, .\label{eq1-4}
\end{equation}

To obtain the tunneling rate, the usual procedure is to
apply the so-called Im-$F$ method. It is based on the
calculation of the free energy $F=(-1/\beta)\log Z$ and
thus of the  partition function $Z$ of an unstable system.
Most conveniently, this is done within the path integral
representation which gives, after tracing out the bath
degrees of freedom exactly, the partition function of the
reduced systems as
\begin{equation}
Z = \int\limits_{q(-\hbar \beta/2)=q(\hbar \beta/2)}
\mathcal{D}q \; e^{-S[q]/\hbar} \label{eq1-4b}
\end{equation}
with the effective Euclidean action functional
$S[q]=S_0[q]+S_{\rm I}[q]$ where
\begin{eqnarray}
S_0[q]= \int_{{-\hbar \beta}/2}^{{\hbar \beta}/2} d\tau
\left[ \frac{M}{2} \; \dot q^2 + V(q)
\right]\label{system_action}
\end{eqnarray}
represents the bare system and
\begin{equation}
S_{ I}[q]= -\frac{1}{2}\int_{{-\hbar \beta}/2}^{{\hbar
\beta}/2} d\tau \int_{{-\hbar \beta}/2}^{{\hbar
\beta}/2} d\tau' \, K(\tau-\tau')\, q(\tau)
q(\tau')\label{eq1-5}
\end{equation}
originates from the coupling to the heat bath (influence
functional). The influence kernel follows from
\begin{equation}
K(\tau)=\int_0^\infty \frac{d\omega}{\pi}\, J(\omega)\,
\frac{{\rm cosh}[\omega(\hbar\beta/2- \tau)]}{{\rm
sinh}(\omega\hbar\beta/2)}\,  \label{eq1-6}
\end{equation}
and is directly related to the analytic continuation to
imaginary times of the force-force autocorrelation function
of the bath $L(t)=\langle \hat{\xi}(t)\hat{\xi}(0)\rangle$
where $\hat{\xi}=\sum_i c_i x_i$, i.e.\
$K(\tau)=L(-i\tau)/\hbar$. Further, due to the relation
(\ref{eq1-4}) the influence kernel can be inferred from
the classical damping kernel \cite{report,weiss}.

The above path integral (\ref{eq1-4b}) sums over all closed
paths in the imaginary time interval
$[-\hbar\beta/2,\hbar\beta/2]$. In case of sufficiently
high potential barriers a semiclassical approximation
applies so that the partition function is dominated by all
periodic minimal action paths and Gaussian fluctuations
around them. It turns out that while fluctuations around a
well minimum of a metastable potential are stable, those
around a barrier top are not. As shown by Langer
\cite{langer}, due to an analytic continuation this gives
rise to an imaginary contribution to the partition
function $Z\approx Z_{\rm w}+ i Z_{\rm b}=Z_{\rm w} (1+i
Z_{\rm b}/Z_{\rm w})$ which, even though it is
exponentially small compared to the well contribution
$Z_{\rm w}$, must be taken into account to finally obtain
the escape rate \cite{langer,caldeira_leggett}
\begin{equation}
\Gamma=-\frac{2}{\hbar} {\rm Im}F=\frac{2}{\hbar\beta}
\frac{Z_{\rm b}}{Z_{\rm w}}\, . \label{eq1-6b}
\end{equation}
For the non-dissipative case and $T=0$ the method is
completely equivalent to the WKB formalism and in other
ranges its results have been verified  by full dynamical
approaches \cite{ankerhold1} (see also
\cite{waxman_leggett}). Its theoretical predictions have
been thoroughly tested in comparison with experimental
data over broad temperatures   ranges and for various
systems \cite{weiss}. The advantage of the method is that
it
 enables us to calculate the decay rate of a metastable
system from a purely thermodynamic quantity without
considering the complicated real time dynamics.

The archetypical form of a metastable potential is
\begin{equation}
V(q)=\frac{M}{2}\omega_0^2 q^2\left(1-\frac{q}{q_0} \right)
\label{potential}
\end{equation}
with a well located around $q=0$ and a barrier top at
$q_{\rm b}=2 q_0/3$. The barrier height is $V_b=2/27 M
\omega_0^2 q_0^2$ and the well frequency $\omega_0$.
Metastability justifying a semiclassical approximation to
the partition function then means that $V_b\gg
\hbar\omega_0, k_{\rm B} T$. For a current-biased
Josephson junction the known tilted washboard potential
for its phase takes locally the above form of a cubic
surface.

Accordingly, in the low temperature domain $Z_{\rm b}$ is
dominated by the so-called bounce orbit, a minimal action
path that runs in the time interval
$[-\hbar\beta/2,\hbar\beta/2]$ through the inverted
barrier potential $-V(q)$ with minimal action $S_b$. The
corresponding tunneling rate turns out to be
\begin{eqnarray}
\Gamma &=&  \sqrt{\frac{S_b}{2 \pi \hbar}}
\sqrt{\frac{D_0}{D'_1}}\ e^{- S_b/\hbar}\label{rate}
\end{eqnarray}
where $D_0$ and $D'_1$ are functional determinants
capturing Gaussian fluctuations around the constant well
orbit $q(\tau)=0$ and the bounce, respectively.  In the
latter one, the contribution from a zero mode direction in
function space is omitted. In absence of a heat bath and
for $T=0$  the above expression leads to
\begin{equation}
\Gamma_0=6 \omega_0\, \sqrt{\frac{6 v}{\pi}}\ {\rm e}^{-36
v/5} \label{nonrate}
\end{equation}
where
\begin{equation}
v=\frac{V_b}{\hbar\omega_0}\label{dbarrier}
\end{equation}
 is the dimensionless barrier height.

\section{Sluggish bath}\label{adiabatic}

While the impact of moderate to fast noise on tunneling
rates has been elucidated analytically and numerically in
the past \cite{weiss}, here, we focus on the low frequency
limit and start with a sluggish bath. This means that the
fastest modes available in the environment are still slow
on the thermal time scale and compared to the system's well
frequency. Specifically, we assume a spectral density
$J_<(\omega)$ with a cut-off frequency $\omega_{c}$, i.e.\
$J_<(\omega)=0, \mbox{for} \ \omega>\omega_{c}$, obeying
$\hbar\beta\omega_{c}\ll 1$ and $\omega_c/\omega_0\ll 1$.
As a consequence, the
influence kernel becomes approximately constant%%%%% approximately
\begin{equation}
K(\tau)\approx \frac{2}{\hbar\beta}\, \int_0^\infty
\frac{d\omega}{\pi} \frac{J_<(\omega)}{\omega}\equiv
\frac{2}{\hbar\beta}\, J_s\ ,\label{eq2-1}
\end{equation}
 and the influence functional simply reads
\begin{equation}
S_{I}[q]\approx \frac{-J_s}{\hbar\beta}\left[\int_{{-\hbar
\beta}/2}^{{\hbar
\beta}/2} d\tau \, q(\tau)\right]^2\, .\label{eq2-2}
\end{equation}
Any bath coupling establishes  correlations with
corresponding  memory times that relate the time evolution
of the system at the present with its past. The time
independent influence kernel of  (\ref{eq2-2}) can be
understood as the limiting case where the frequencies of
the environmental degrees of freedom are are so low, that
the critical correlation time ranges over the whole
interval $[-\hbar\beta/2,\hbar\beta/2]$.

The  sluggish bath can now most conveniently be treated by
exploiting a Hubbard-Stratonovich transformation and
introducing an auxiliary variable. Accordingly, the
partition function is obtained as
\begin{equation}
Z(\sigma_x) = \sqrt{\frac{1}{2\pi \sigma_x}}
\int_{-\infty}^{\infty} \mit d x \exp \left(-\frac{x^2}{2
\sigma_x} \right) Z(x) ,\label{zustandssumme}
\end{equation}
where $Z(x)$ is the path integral
\begin{equation}
Z(x)=\int \mathcal{D}q \; e^{-S_x[q]/\hbar}\ .
\end{equation}
The action functional $S_x[q]=S_0[q]+S_{x,I}[q]$ with
\begin{equation}
S_{x,I}[q] = x \int_{-\hbar \beta/2}^{\hbar \beta/2}d\tau\,
q(\tau)  \label{gauss_action}
\end{equation}
can be seen as a linear coupling between the "centroid" of
the imaginary time orbit and a constant "external force".
The total partition function is thus represented as an
average over individual partition functions where the
contribution of each one is weighted according to a
Gaussian distribution with width $\sigma_x=2 J_s/\beta$.

Now, by a proper coordinate shift for a fixed value of the
auxiliary variable $q = \tilde{q} + Q(x)$, the $S_{x,I}$
contribution in the effective action is absorbed in a
potential contribution of an effective bare system. While
the measure of the path integral remains unchanged, the
parameters of the shifted potential become $x$-dependent,
namely,
\begin{equation}
\tilde{V}(q,x) = \frac{M}{2}\; \tilde{\omega}_0(x)^2 q^2
\left( 1-\frac{q}{\tilde{q}_0(x)} \right), \label{pot x}
\end{equation}
where $\tilde q$ and $\tilde{\omega}_0$ are defined by
demanding that ${xq + V(q)} = \tilde{V}(\tilde{q},x) +
V(Q(x)) + x Q(x)$. With the abbreviation
\begin{equation}
\alpha(x) = \sqrt{1+\frac{6x}{M \omega_0^2 q_0}}
\end{equation}
 one
obtains the coordinate shift and the modified parameters as
\begin{eqnarray}
Q(x) &=& \frac{q_0}{3}\left[1- \alpha(x)
\right]\label{shift}
\\ \tilde{\omega}_0(x) &=& \omega_0 \sqrt{\alpha(x)}
\label{omega_schl} \\ \tilde{q}_0(x) &=& q_0 \alpha(x)
\label{q_schl}.
\end{eqnarray}
The partition function for fixed $x$ now takes the form
\begin{eqnarray}
Z(x) = e^{-\beta [V(Q(x))+x Q(x)]} \oint \mathcal{D}
\tilde{q} \; e^{-\frac{1}{\hbar}\tilde{S}_0[\tilde{q};x]}
,\label{zustandssumme1}
\end{eqnarray}
where $\tilde{S}_0[\tilde{q};x]$ denotes the action of the
bare system (\ref{system_action}) with the $x$ dependent
potential $\tilde V$ from (\ref{pot x}).

Now, in line with the semiclassical approximation, for
sufficiently small width $\sigma_x$ (details see below)
the average over the auxiliary variable can be treated
together with the fluctuations around the bounce
$\tilde{q}=\tilde{q}_B+\delta\tilde{q}$ in steepest decent
approximation : The full action is expanded around the
bounce $q_b$ for $x=0$ up to terms of second order. This
results in
\begin{eqnarray}
\tilde{S}_0[\tilde{q};x] &=& S_b(x=0) +
4\frac{q_0}{\omega_0} x + \frac{3}{M
\omega_0^3}x^2  \\
\nonumber && \hspace{-1.5cm} + \frac{1}{2} \int \limits_{{-\hbar
\beta}/2}^{{\hbar \beta}/2} d\tau \; \delta \tilde{q}
\;\frac{M}{2} \left(- \partial_{\tau}^2 +
\frac{\tilde{V}''(\tilde{q}_{B,x}, x=0)}{M} \right) \delta
\tilde{q} \nonumber \\ &+& O(x^3,\delta\tilde{q}^3)
 .\label{term1} \nonumber
\end{eqnarray}
where $S_b$ is the minimal action corresponding to $q_b$.
In this second order approximation the last term in
(\ref{term1}) is independent of $x$ and the remaining path
integral over the fluctuations yields the known functional
determinant for $x=0$. The total rate then factorizes: The
$x=0$-part coincides with the non-dissipative tunneling
rate $\Gamma(\beta)$ in a cubic potential;
 the remaining
$x$-dependent terms in  (\ref{term1}) enter into the
average over the Gaussian distribution.  Accordingly, the
total rate reads
\begin{equation}
\Gamma_s(j,\theta)= \Gamma(\theta) \; I_s(j,\theta) \, ,
\label{rateslow}
\end{equation}
where  we introduced dimensionless quantities
${\theta=2\pi/(\omega_0\hbar\beta)}$ and $j =2 J_s/(M
\omega_0^2)$, and the correction factor
\begin{equation}
I_s(j,\theta) = \sqrt{\frac{1-j}{1-j+\frac{6 j \theta}{2
\pi}} } \;\exp \left[\frac{108 \frac{j\theta}{2 \pi} v }{
(1-j+\frac{6 j\theta}{2 \pi} )} \right]\, . \label{ifactor}
\end{equation}

The correction factor provides a restriction on $\sigma_x$
which in turn defines the range where
 the steepest descent approximation for the $x$-average is
 justified. From $j-6 j \theta/2\pi\ll 1$ one finds
 \begin{equation}
 \frac{J_s}{M\omega_0^2}\ll 1\ .\label{cond}
 \end{equation}
 For example, in case of low frequency Drude friction
 $J_<(\omega)=M\gamma\omega\,
 \omega_c^2/(\omega^2+\omega_c^2)$ this leads to
 $\gamma\omega_c/\omega_0^2\ll 2$ which due to $\omega_c/\omega_0\ll 1$
 allows even for stronger friction. Upon further inspection, one sees that
sufficiently below the crossover temperature $\theta\ll 1$
the
   full rate (\ref{rateslow}) displays a dominating linear
 temperature increase due to its correction factor, while
 the bare rate approaches its zero temperature limit
 $\Gamma_0$  (\ref{nonrate})
 since deviations become exponentially small in
$2\pi/\theta$. This linear enhancement in $T$ is a specific
feature of a sluggish bath $\omega_c\hbar\beta\ll 1$. We
note that thus in (\ref{ifactor}) the limit $T\to 0$ cannot
be performed for fixed $\omega_c$, but must include also
$\omega_c\to 0$. Accordingly, zero-point fluctuations of
the bath modes, which lead to the well-known counter term
and renormalize the bare  potential \cite{weiss,martinis},
do not appear here.
%------------------------- figure: staticrate ------------------------------
\begin{figure}
%\vskip-2.0cm
\includegraphics[width=85mm]{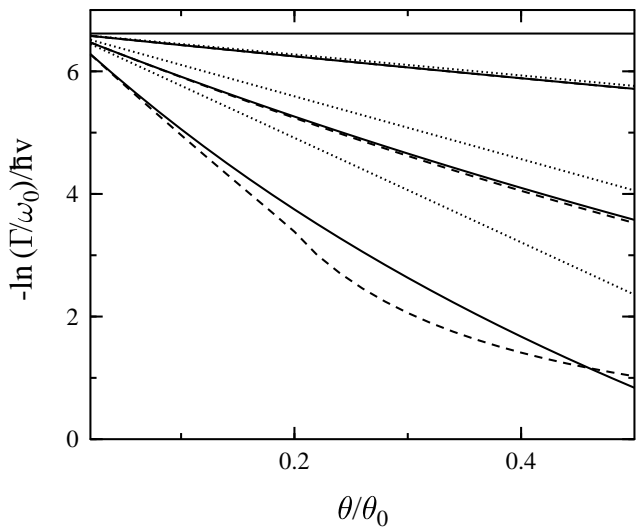}
%\vskip-5.0cm
\caption{MQT rates in presence of a sluggish
bath vs. temperature for various values of the coupling
constant $j=0.0, 0.1, 0.3, 0.5$ (from top to
bottom). Shown are results
according to (\ref{rateslow}) (solid), numerical results (dashed) and perturbative
results according to (\ref{staticper}) (dotted).}
\label{figure_static}
\end{figure}

An alternative procedure to calculate the tunneling in
the limit of very weak dissipation and temperatures $T\ll
T_0$  relies on a perturbative treatment around the
non-dissipative bounce for $T=0$. The first order
correction to the bare bounce action $S_b$ is determined
by the influence functional $S_{ I}[q]$ in (\ref{eq2-2})
evaluated along the $T=0$-bounce $q_b$. Here, we obtain
\begin{eqnarray}
 S_{ I}[q_b]/\hbar = -108\,  \frac{j\,\theta\, v}{2\pi}
 \label{staticper}
\end{eqnarray}
 in accordance with the result obtained above in (\ref{rateslow})
 up to terms linear in $j$. Hence, the expression (\ref{ifactor})
 applies also to larger values for $j$ beyond the validity of the
 first order perturbation theory (see below).
In this context we mention that the same result can be
obtained by considering fluctuations in a control
parameter of the rate, e.g.\ the external bias current for
Josephson junctions, and then taking the average over the
change of the dominant exponential factor with  an
appropriate Gaussian distribution \cite{martinis}.

In order to illustrate the accuracy of the expression
(\ref{rateslow}), we evaluated
the rate numerically
according to the general formula (\ref{rate}) in 
the case of weaker dissipation
and a sluggish bath (see fig.~\ref{figure_static}). The
bounce is gained from its Fourier coefficients
$q_b(\tau)=\sum Q_n \exp(i\nu_n\tau)$ according to the
equation of motion in the inverted potential (see
\cite{olschowski} for details). For the comparison, we
further approximated $\Gamma(\theta) \approx \Gamma(0)$ in
 (\ref{rateslow}) which is correct up to exponentially
small corrections for low temperatures. The agreement with
the analytical result is excellent even for
$j=0.5$ and sufficiently low temperatures, while
the perturbative result displays negligible deviations
from the numerical data only for $j\leq 0.1$. Obviously,
the dissipative rate is always enhanced roughly linearly
compared to the bare one due to the impact of the sluggish
bath fluctuations.

\begin{figure}
%\vskip-2cm
\includegraphics[width=85mm]{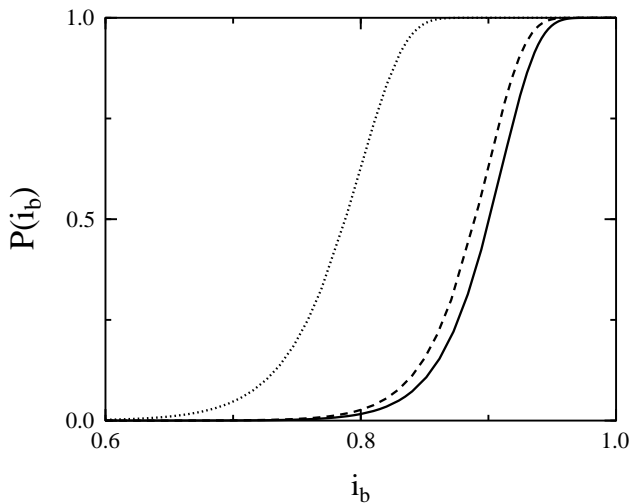}
%\vskip-5cm
\caption{Probability for escape vs.\ bias current for a
Josephson junction with dimensionless barrier height $v=3$
and various values of the coupling to a sluggish bath
$j=0.0 \ (\mbox{solid}), 0.1\ \mbox{(dashed)}, 0.3\
\mbox{(dotted)}$. } \label{slow-s-curves}
\end{figure}

In experiments on Josephson junction circuits typically
the escape rate is not measured directly, but rather the
probability for escape as the height $i_b$ of an external
bias current pulse varies, while keeping the width $\Delta
t$ in time of the pulse constant, i.e.,
\begin{equation}
P(i_b)=1-{\rm e}^{-\Gamma(i_b)\, \Delta t}\, .
\label{probi}
\end{equation}
The parameters of the cubic potential (\ref{potential})
depend for $i_b$ close to the critical current $i_0$ of
the junction on the bias current via
$\omega(i_b)=\omega(i_b=0) [1- (i_b/i_0)^2]^{1/4}$ and
$V_b(i_b)=V_b(i_b=0) (1-i_b/i_0)^{3/2}$. In
fig.~\ref{slow-s-curves} the corresponding "s-curves" are
depicted for various strengths of the coupling. Since in
presence of slow bath modes the rate is larger than the
bare one, the s-curves are shifted towards  smaller values
of the bias current and exhibit a smaller slope.
% ---------------------figure: slow-s-curves -------------------%%

\section{Sluggish Bath and $1/f$-damping}\label{dynamic}

Having analyzed the influence of the coupling to sluggish
bath modes we now turn to the case where in addition to a
pronounced low frequency part a tail of slow to moderate
frequency modes are present. In particular, we consider
dynamical fluctuations the power spectrum of which display
a $1/f$ characteristic.

\subsection{Spectral density and influence kernel}\label{kernel}

 In present experiments on
superconducting qubits $1/f$ noise has been found to be the
most dominant (and most annoying) source of noise
\cite{zorinahlers}. While in principle one could start
with a truly microscopic model as in
\cite{ankerepl,paladino}, here, we proceed with a somewhat
simplified procedure and effectively mimic the $1/f$ noise
by a proper spectral density of a heat bath.  We write
\begin{equation}
J(\omega)=J_<(\omega)+J_{1/f}(\omega)\label{spectruma}
\end{equation}
where  $J_<$ with $J_<(\omega)=0$ for $\omega>\omega_{c}$
comprises all sluggish environmental modes  and $J_{1/f}$
with $J_{1/f}(\omega)=0$ for $\omega<\omega_-$ and
$\omega>\omega_+$ contains its dynamical modes. For
convenience, we introduced here in addition to $\omega_c$
the lower cut-off $\omega_-$ with $\omega_c\leq \omega_-$
to have better control over the two domains of the
spectrum. Clearly, the separation  into a sluggish part
and a dynamical one is somewhat arbitrary and depends on
the specific experimental situation,  i.e.\ on the relevant
time scales involved and the measurement protocol imposed.
We will see at the end, however, that our main findings
are independent
 of these latter cut-off frequencies as long as they lie in a
range $\omega_c, \omega_-\ll \omega_0$. The upper cut-off
$\omega_+$ is taken to be only somewhat smaller than
$\omega_0$ such that  $2\pi/\omega_+\hbar\beta$ is still
sufficiently larger than 1. For instance, for typical
experimental values $\omega_0\simeq 50$ GHz and $T\simeq
20$ mK corresponding to $\omega_0\hbar\beta\approx 20$,
one may assume $\omega_+$ to lie in the GHz range as well,
i.e.\ $\omega_+\simeq 1$ GHz, leading to
$2\pi/\omega_+\hbar\beta\approx 15$.

Now, the power spectrum of bath fluctuations follows from
the symmetrized force-force auto-correlation function of
$\hat{\xi}=\sum c_{\alpha} x_{\alpha}$ as
\begin{equation}
\tilde{S}_{\xi}(\omega) =
\frac{1}{2\pi}\int_{-\infty}^{+\infty} e^{i \omega t}
\langle \hat{\xi}(t) \hat{\xi}(0)\rangle_{\beta} dt\,
.\label{power_spectrum}
\end{equation}
Experimentally,   background charge fluctuations in
mesoscopic circuits have shown to give rise to a behavior
$\tilde{S}_{\xi}(\omega)= s_0/\omega$ over a broad
frequency range ($1/f$ noise). According to
(\ref{spectruma}) the spectral density $J_{1/f}$ of the
{\em dynamical} modes can then be calculated using the
fluctuation-dissipation-theorem
\begin{equation}
\frac{s_0}{\omega}\ \Theta(\omega - \omega_{-})
\Theta(\omega_{+} - \omega)= \hbar \coth (\omega \hbar
\beta) J_{1/f}(\omega) .\label{spectrum}
\end{equation}

As a consequence of (\ref{spectruma}) the influence kernel
$K(\tau)$ splits into two contributions where the usual
procedure is to split off from the second one a part local
in time, i.e.,
\begin{equation}
K(\tau)=\frac{2 J_s}{\hbar\beta}-k(\tau)+ J_f
:\delta(\tau):\label{kone}
\end{equation}
where
\begin{equation}
J_f=\int_0^\infty\frac{d\omega}{\pi}\, \frac{2\,
J_{1/f}(\omega)}{\omega} \label{ktwo}
\end{equation}
and
\begin{equation}
k(\tau)=\frac{M}{\hbar\beta}\sum_{n=-\infty}^\infty
|\nu_n|\, \hat{\gamma}(|\nu_n|)\, {\rm e}^{i \nu_n\tau}\ .
\label{kthree}
\end{equation}
Here, $:\delta(\tau):$ is the periodically continued
$\delta$ function with period $\hbar\beta$ and $\nu_n=2 \pi
n/\hbar\beta$ denote the Matsubara frequencies.

The Fourier coefficients of the dynamical part of the
influence kernel are  obtained from the spectral density
via the relation (\ref{eq1-3}). This way, for a  spectral
density with dynamical modes one finds from
(\ref{spectrum}) $J_{1/f}(\omega)=\kappa_{1/f}(\omega)\, M
{\Theta(\omega-\omega_-)}{\Theta(\omega_+-\omega)}$ where
in the relevant frequency domain $\kappa_{1/f}(\omega)=s_0
\,{\rm tanh}(\omega\hbar\beta)/(\hbar\omega M)\approx
\kappa$ is basically constant in frequency and taken to be
independent of temperature as well. Consequently, the
amplitude of the power spectrum $s_0$ increases  linearly
with temperature $T$ in agreement with an electron trapping
mechanism for $1/f$ noise \cite{dutta}. Eventually, one
arrives at
\begin{equation}
\hat{\gamma}(|\nu_n|)=\frac{\kappa}{|\nu_n|\pi}  \log
\left({\frac{1+\nu_n^2/\omega_{-}^2}{1+\nu_n^2/\omega_{+}^2}}
\;\right)\ .\label{hat}
\end{equation}
 In particular, $|\nu_n|\hat{\gamma}(|\nu_n|)=0$ for $n=0$, while for $n\neq 0$
due to $\nu_n/\omega_-\gg \nu_n/\omega_+\gg 1$ one expands
$|\nu_n|\hat{\gamma}(|\nu_n|)\approx (2\kappa/\pi)
 \log(\omega_+/\omega_-)-(\kappa/\pi)
 (\omega_+^2-\omega_-^2)/\nu_n^2$.
The constant factor $(2\kappa/\pi)
 \log(\omega_+/\omega_-)$ leads in the time domain to a
contribution proportional to $  [
:\delta(\tau):-1/\hbar\beta]$ where the part containing the
$\delta$ function exactly cancels in (\ref{kone}) the
contribution proportional to $J_f$. In turn, this means
that a spectral density corresponding to $1/f$ noise does
{\em not} generate a contribution in the influence
functional local in time and renormalizing the potential.

Now, the influence functional (\ref{eq1-5}) takes the form
\begin{eqnarray}
S_I[q]&=&-\frac{\tilde{J}_s}{\hbar\beta}
\left(\int_{-\hbar\beta/2}^{\hbar\beta/2} d\tau
q(\tau)\right)^2  \\ \nonumber &+& \frac{1}{2}
\int_{-\hbar\beta/2}^{\hbar\beta/2}d\tau
\int_{-\hbar\beta/2}^{\hbar\beta/2} d\tau'\ q(\tau)
\tilde{k}(\tau-\tau') q(\tau')\label{1overfaction}
\end{eqnarray}
with
\begin{equation}
\tilde{J}_s=J_s+(M
\kappa/\pi)\log(\omega_+/\omega_-)\label{jtilde}
\end{equation}
 being an effective sluggish bath coupling and a non-local damping kernel
\begin{equation}
\tilde{k}(\tau)=\zeta\, \frac{
M}{\hbar\beta} \sideset{}{'} \sum_{n=-\infty}^{\infty} \frac{1}{\nu_n^2}\
{\rm e}^{i \nu_n \tau}\, ,\label{ktilde}
\end{equation}
where in the sum the $n=0$ term is omitted and
\begin{equation}
\zeta=\kappa (\omega_+^2-\omega_-^2)/\pi\, \approx
\kappa\omega_+^2/\pi .\label{zeta}
\end{equation}
In fact, the above Fourier series can be summed up exactly
and yields $\tilde{k}(\tau)=\sum_{l=-\infty}^\infty
\tilde{k}_l(\tau)$ with
\begin{eqnarray}
\tilde{k}_l(\tau)&=& \frac{M\zeta}{2\hbar\beta}
\left[\tau-(2l+1)
\frac{\hbar\beta}{2}\right]^2 \\ \nonumber &-& \frac{M\zeta\hbar\beta}{24}
\quad \; \quad \mathrm{for} \:\tau\in \big(l\hbar\beta,
(l+1)\hbar\beta\big)\, .\nonumber \label{finalk}
\end{eqnarray}
This way, the influence of the sluggish modes is described
by one effective parameter $\tilde{J}_s$ independent of
the specific cut-off frequencies $\omega_c$ and $\omega_-$,
and the only relevant frequency scale of the bath is
$\omega_+$.

\subsection{Tunneling rate}\label{1/ftunneling}

With the influence kernel at hand, the tunneling rate can
now be determined.  The sluggish modes leading to the
static contribution in (\ref{1overfaction}) are treated
analogous to the procedure described in the previous
Sect.~\ref{adiabatic}. Note that the corresponding
constant coordinate shift does not change the form of  the
dynamical part since $\int_0^{\hbar\beta} d\tau
\tilde{k}(\tau)=0$. The effective action is obtained as
$\tilde{S}[\tilde{q};x]=\tilde{S}_0[\tilde{q};x]+\tilde{S}_I[\tilde{q}]$
with
\begin{equation}
\tilde{S}_I[q]=\frac{1}{2}\int_{{-\hbar
\beta}/2}^{{\hbar
\beta}/2} d\tau \int_{{-\hbar \beta}/2}^{{\hbar
\beta}/2} d\tau' \, \tilde{k}(\tau-\tau')\, \tilde{q}(\tau)
\tilde{q}(\tau')
\end{equation}
with the $x$-dependent potential given in (\ref{pot x}).

For the case of weak dissipation and low temperatures
damping due to dynamical modes is treated perturbatively
to first order in the dimensionless coupling
$\tilde{\kappa}=\kappa/\omega_0^2$ by substituting the
undamped $T=0$-bounce into the influence functional $S_I$.
Clearly, due to the $x$-dependence of the bare action,
this orbit becomes also $x$-dependent and reads
\begin{equation}
\tilde{q}_{b,x}(\tau) = \frac{\tilde{q}_0(x)}{\cosh ^2
(\tilde{\omega}_0(x) \tau/2)} \label{bouncex}.
\end{equation}
Then, for the total partition function one arrives at
\begin{eqnarray}
Z(\sigma_x,\alpha,\beta) &=& \sqrt{\frac{1}{2 \pi \sigma_x}}
\int_{-\infty}^{\infty} d x \; e^{-\frac{x^2}{2 \sigma_x } } \nonumber \\
&\times& e^{-\frac{1}{\hbar} S_{1/f,\rm  dyn}(x)} \;Z(x),
\label{zustandssumme_over1}
\end{eqnarray}
where $Z(x)$ is specified in  (\ref{zustandssumme1}) and
$S_{1/f,\rm  dyn}(x)=\tilde{S}_I[\tilde{q}_{b,x}]$. Again
the $x$-integration can be carried out by expanding the
exponent up to second order in $x$. The resulting rate
expression   is given by
\begin{equation}
\Gamma_{s,1/f}(\tilde{j},\tilde{\kappa},\theta) =
\Gamma(\theta) \ I_{s,1/f}(\tilde{j},\tilde{\kappa},\theta)
\label{ratealpha}
\end{equation}
with the dimensionless temperature
$\theta=2\pi/\omega_0\hbar\beta$ and the dimensionless
couplings $\tilde{j}= 2\tilde{J}_s/M\omega_0^2$ and $
 \tilde{\kappa}=\kappa/\omega_0^2$. The correction factor
 is now obtained as
\begin{eqnarray}
I_{s,1/f}(\tilde{j},\tilde{\kappa},\theta) &=&
\sqrt{\frac{1-\tilde{j}} {1-\tilde{j}+6\tilde{j}\theta /
2\pi}} \,\exp\left[-S_{1/f,\rm
dyn}(0)/ \hbar\right] \nonumber \\ && \hspace{-1cm}\times  \exp\left[\frac{27
\theta \tilde{j} v}{\pi}\ \frac{2 +  \omega_0\,
S'_{1/f,\rm  dyn}(0)/q_0 }{ 1-\tilde{j}+6\tilde{j}\theta /
2\pi} \right] .\label{ifactorover1}
\end{eqnarray}
Here $S'_{1/f,\rm  dyn}$ denotes the derivative of the
damping term with respect to $x$ and  is of first order in
$\tilde{\kappa}$ as well as $S_{1/f, \rm dyn}$.
Apparently, the influence of the environment leads to two
competing effects: The static modes give rise to a rate
enhancement, while the contribution of the dynamical ones
depresses the rate. To get further insight, we thus
evaluate  the action $S_{1/f, \rm dyn }(0)$ analytically
by exploiting the results in (\ref{finalk}) and
(\ref{bouncex}), and obtain
\begin{equation}
S_{1/f,\rm  dyn}(0)/\hbar= \frac{M\tilde{\kappa} q_0^2
\omega_+^2 }{\pi\hbar\omega_0}
\left(\frac{a_1}{\omega_0\hbar\beta}-2
a_2+\frac{2 \omega_0\hbar\beta}{3}\right)\label{analyticalaction}
\end{equation}
with constants $a_1=26.319..., a_2=4.00...$. In the low
temperature range $\omega_0\hbar\beta\gg 1$ this reduces to
\begin{equation}
S_{1/f,\rm  dyn}(0)/\hbar\approx  v \frac{9
\tilde{\kappa}}{\pi }\, \, \frac{\omega_+}{\omega_0}\, %%%%% canceled the 2
\omega_+\hbar\beta \label{approxactin}
\end{equation}
where we used (\ref{zeta}). An analogue calculation for the derivative of the
damping term yields
\begin{eqnarray}
S'_{1/f,\mathrm{dyn}}(0) &\approx& \frac{6}{M \omega_0^2
q_0} S_{1/f,\rm{dyn}}(0)
 \\ \nonumber && \hspace{-2.0cm}- \frac{3}{M \omega^2_0 q_0} \; \frac{27 \hbar v \tilde{\kappa}
\omega_+^2}{2 \pi \omega_0^2}
\left(\frac{b_1}{\omega_0\hbar\beta}-2
b_2+\frac{\omega_0\hbar\beta}{3} b_3\right) \label{approxactder}
\end{eqnarray}
with the constants $b_1=52.6379...$, $b_2=6.00$ and ${b_3=2.00}$. %%%%% the derivative

We can now compare the leading order terms in the two
exponents of the correction factor (\ref{ifactorover1}).
With respect to the static part,   it is sufficient to
consider only the dominant term $54 \theta \tilde{j}
v/\pi$ and there to focus on the contribution stemming
from $J_{1/f}$, namely, $S_{1/f, \rm stat}= 108 \theta v
\kappa \log(\omega_+/\omega_-)/(\pi^2\omega_0^2)$.
Accordingly,
\begin{equation}
\frac{S_{1/f, \rm stat}}{S_{1/f, \rm dyn}}=\frac{48
\log(\omega_+/\omega_-)}{(\omega_+\hbar\beta)^2}\gg 1\, ,
\label{relation}
\end{equation}
which reveals that the dynamical contribution of the
environment is basically negligible compared to the static
one.  Hence, the system-bath coupling parameters obey the
relation $\tilde{j}/\tilde{\kappa}\gg 2
(\omega_+\hbar\beta/2 \pi)^2$. For the tunneling process a
low to moderate frequency bath with a $1/f$ characteristic
can thus be treated as static and always leads to a rate
enhancement which  roughly grows linearly with
temperature. The physical reason for this is simply that
the environment has only a time period of the order of
$1/\omega_0$ to probe the system while tunneling. On this
time scale the dominant part of the bath modes is
basically frozen, while the contribution of the dynamical
ones remains small for a $1/f$ spectrum. A changeover from
a short to a  long time behavior cannot be observed in
contrast to the decay of coherent superpositions in
superconducting qubits
\cite{shnirman,clarke,averin,paladino2}.

\begin{figure}
%\vskip-2.0cm
\includegraphics[width=85mm]{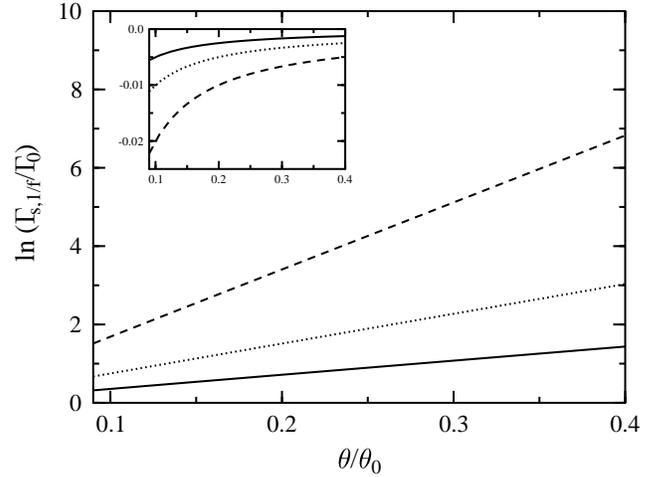}
%\vskip-5.0cm
\caption{Rate enhancement due to a sluggish bath with $1/f$
noise vs.\ temperature. The couplings constants are
$\tilde{j}=\tilde{\kappa}=0.05$ (solid), 0.1 (dotted), 0.2
(dashed), and $\omega_+/\omega_0=1/50$. The inset shows
the suppression only due to dynamical modes
($\tilde{j}=0$), but with the same values for
$\tilde{\kappa}$} \label{figure-enhancement}
\end{figure}

In order to illustrate these findings we show in
fig.~\ref{figure-enhancement} the ratio
$\Gamma_{s,1/f}(\theta)/\Gamma_0$ for varying temperature
and different damping strength $\tilde{j}=\tilde{\kappa}$.
The analytical expression obtained in  (\ref{ratealpha})
coincides within this range
 with the numerical data. As seen in the inset, the
 influence of the dynamical modes to suppress the MQT rate
 is very small and only increases slightly at very low
 temperatures. Note again, that for fixed $\omega_c$ the
 limit ${T\to 0}$ cannot be reached. Typical
 temperatures in experiments, however, lie in the range about 20 mK
 corresponding to $\theta/\theta_0\simeq 0.1\ldots 0.3$.
%-------------------- figure: 1overbath enhancement ---------------------------

\section{Quantronium qubit}\label{quantronium}

The "quantronium" is a solid state based qubit setup
consisting of a  Cooper pair box whose Josephson junction
is split in two small junctions with Josephson energy
$E_J$, delimiting an island with capacitance $C$ and
charging energy $E_C=(2 e)^2/2 C$ \cite{denis}. The two
lowest lying states of the box with energies $E_{0}$ and
$E_1$ encode a qubit and in the circuit give rise to loop
currents of opposite direction. In parallel with  the box
is a third big Josephson junction (with $E_J'\gg E_J$ and
$E_C'\ll E_C$) which serves as a detector. For the
read-out this big junction is biased by an external
current pulse so that the total bias current seen by the
junction depends on the state of the qubit. The
measurement consists of adiabatically driving the big
junction into the regime where MQT sets in. Due to the
exponential sensitivity of the tunneling rate on the total
bias current, the two qubits states can efficiently be
discriminated. In the quantronium two dominant sources of
noise influence the qubit, namely, charge noise and phase
noise. The latter one is mainly due to noise in the big
read-out junction and there, originates e.g.\ from
fluctuations of the external bias current. In the sequel,
we are interested in this low frequency noise affecting
the phase of the big junction during the tunneling process.

The analysis of the device becomes particularly
transparent in the charging regime ($E_C\gg E_J$) where
the two qubit states are determined by superpositions of
zero or one excess Cooper pairs on the island. Then,
 the Hamiltonian of the box-subsystem can be written in terms of
Pauli matrices and by measuring all energies in units of
$E_J'$ the total dimensionless Hamiltonian of the circuit
reads
\begin{equation}
h = e_C\, \sigma_z - e_J\, \cos \left(\frac{\theta +
\phi}{2} \right) \sigma_x + \frac{P^2_{\theta}}{2M}-
\cos(\theta) - i_b \theta \label{qubit_hamilton}
\end{equation}
with the dimensionless parameters
$e_C=(E_C/E_J'){(N_g-1/2)}$, $e_J=E_J/E_J'$, and $i_b=\hbar
I_b/2 e E_J'$. Here, $N_g$ is the reduced gate charge and
$I_b$ denotes the external bias current.
 $\phi$ represents  the reduced magnetic flux in units of
 $\hbar/2e$ and $M=C' E_J'/4 e^2$ is the mass of the
 artificial particle of the read-out junction with momentum $P$
 and conjugate phase
 $\theta$ the dynamics of which takes place in a tilted
 washboard potential $- \cos(\theta) - i_b
\theta$. For further details we refer to
 \cite{denis,ankerhold}. Due to the coupling between the two subsystems, the qubit and the big
 junction, the dynamics of $\theta$ can be seen as that of
 a particle with two internal states.

 Now, we consider the situation when the qubit is initially
 ($i_b=0$) prepared in its ground state. By rising the
 bias current $i_b>0$ adiabatically the phase $\theta$ can
 be seen as a classical variable with negligible kinetic
 energy due to the large capacitance $C'$. Depending on the
 spin state, the artificial particle then evolves on
 adiabatic potential surfaces $\lambda_\pm(\theta)$
 obtained by simply diagonalizing $h$ in spin space for
 $M\to\infty$ \cite{ankerhold}. When $i_b$ is close to 1, however,  the particle may
 tunnel out of the potential well. In
 this case the Hamiltonian is most conveniently represented
 in the spin basis at the minimum $\theta_{\rm min}$ of the lower
 surface $\lambda_-(\theta)$, i.e.,
\begin{equation} H = \left(
\begin{array}{ll}
\frac{p^2_{\theta}}{2M} + V_+(\theta)    & \Delta(\theta) \\
\Delta(\theta)            &\frac{p^2_{\theta}}{2M} +
V_-(\theta)\, .
\end{array}
\right) \label{dia_hamilton}
\end{equation}
Here,  the diabatic potential surfaces read
\begin{equation}
V_{\pm} = - \cos(\theta) - i_b \theta \pm \left(
\sqrt{e_C^2+ V_0^2} +
\frac{V_0[V(\theta)-V_0]}{\sqrt{e_C^2+ V_0^2}}\right)
\end{equation}
and the off diagonal elements
\begin{equation}
\Delta(\theta) = \frac{e_C [V_0-V(\theta)]}{\sqrt{e_C^2+
V_0^2}}
\end{equation}
where $V(\theta)=e_J \cos(\frac{\theta + \phi}{2})$ and
$V_0=V(\theta_{\rm min})$. By construction,  for
$\theta=\theta_{\rm min}$ the off-diagonal elements vanish.
Further, one approximates these diabatic surfaces in the
well-barrier range by a cubic expansion around the well
minimum $\theta_{\rm min}$
\begin{equation}
V_{\pm}(q)= \frac{M
\Omega_{\pm}^2}{2}\;(q-q_{\pm})^2[1-(q-q_{\pm})/q_{0,
{\pm}}] + \delta_{\pm,+}\, \Delta V_{\rm min},
\end{equation}
where $q=\theta- \theta_{\rm min}$ is measured relatively
to the minimum, $\delta_{\mu,\nu}$ denotes the Kronecker
symbol, and $\Delta V_{\rm min}=V_+(\theta_{\rm
min})-V_-(\theta_{\rm min})$.
%------------------------- figure: diabaticpotential  ------------------------
\begin{figure}
\includegraphics[width=85mm]{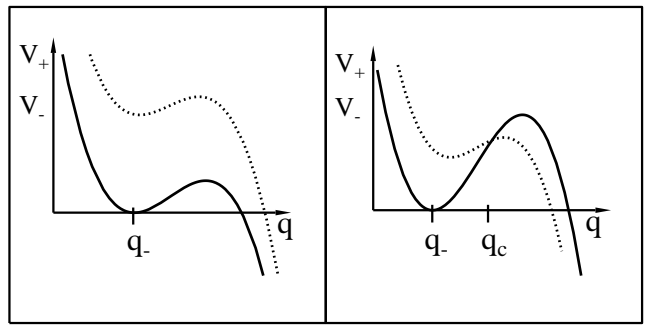}
\caption{Diabatic potential surfaces $V_+$
(dashed) and $V_-$ (solid) outside (left) and inside
(right) the range where Zener flip tunneling occurs.}
\label{figure-diabatic}
\end{figure}

Now, as long as $V_\pm$ are sufficiently separated from
each other everywhere in the barrier range, i.e.
$|V_--V_+|\gg \Delta$, the spin degree of freedom is
essentially frozen and the particle escapes  through $V_-$
via standard MQT (see fig.~\ref{figure-diabatic}, left).
The influence of noise for this scenario is thus captured
by the findings of the previous sections.

There is an additional domain, however, which gives rise
to an interesting MQT phenomenon \cite{ankerhold,zener2}.
Namely, when the two diabatic surfaces cross each other in
the barrier range (see fig.~\ref{figure-diabatic}, right),
the particle's spin may flip during the tunneling since
this may enhance the probability for escape substantially.
The standard MQT approach has then to be extended to
include spin flips so that the tunneling rate follows, in
case of vanishing dissipation, from the imaginary part of
the partition function
\begin{eqnarray}
Z &=& \int \mathcal{D}q \; e^{- S_{-}[q]} \; \Bigg\{ 1 +
\sum_{n=1}^{\infty} \int
\limits_{-\hbar\beta/2}^{\hbar\beta/2} d\tau_{2n} \cdots
\int
\limits_{-\hbar\beta/2}^{\tau_2} d\tau_1 \nonumber \\
&\times& \Delta(q(\tau_{2n})) \cdots \Delta(q(\tau_1)) \nonumber \\
&\times& \exp\left[\sum_{k=1}^{n} \,
\int \limits_{\tau_{2k-1}}^{\tau_{2k}} d\tau (V_{-}(q) -
V_{+}(q))\right]\Bigg\},\label{summeflip}
\end{eqnarray}
where $S_-$ denotes the bare action on the surface $V_-$.
The corresponding MQT rate can be cast into
\begin{equation}
\Gamma_{\rm tot}=\Gamma_b+f_{\rm flip}(\Delta_c)\,
\exp(-S_{\rm flip}/\hbar)
\end{equation}
where $\Gamma_b$ denotes the standard MQT rate without
spin flip, while the second term captures contributions
due to flips. The prefactor $f_{\rm flip}(\Delta_c)$ with
$\Delta_c$ the coupling at the crossing point of $V_\pm$
is related to the probability for a spin flip to occur and
$S_{\rm flip}$ is the action along the flip-bounce. It
turns out that a theory based on the non-dissipative
partition function (\ref{summeflip}) provides rates which
are already in good agreement with experimental data
\cite{zener2}. Nevertheless, a deeper understanding of the
impact of low frequency noise on the Zener-flip-tunneling
is clearly needed.

\begin{figure}
%\vskip-2.5cm
\includegraphics[width=85mm]{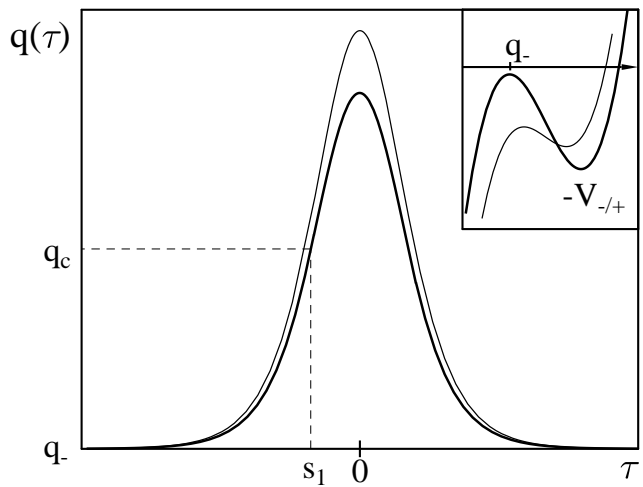}
%\vskip-6.5cm
\caption{Ordinary bounce (thin) and flip bounce
trajectory (thick). The inset displays the inverted
diabatic potentials ($V_+$ thin, $V_-$ thick) intersecting
at $q_c$.} \label{figure-bounce}
\end{figure}

It was found in \cite{ankerhold} that when the particle
traverses the intersection range of $V_\pm$ sufficiently
fast the flip contribution is dominated by two spin-flips.
Further, in case of weak friction the most profound effect
of dissipation on the rate is provided by the action
factor. Hence, we focus on the two-flip bounce and its
action at low temperatures. In generalization of
(\ref{eq1-4b}), the dissipative flip-bounce action takes
the form
\begin{eqnarray}
S_{\rm flip}[q;s_1,s_2] &=& \nonumber \\ && \hspace{-2.5cm} \int
\limits_{-\hbar\beta/2}^{\hbar\beta/2} d\tau \left\{
\frac{M}{2} \; \dot q^2 + V_{-}(q)+ h_{s_1,s_2}(\tau) \,
[V_+(q)-V_-(q)]\right\} \nonumber
\\ &-& \frac{1}{2} \int\limits_{-\hbar\beta/2}^{\hbar\beta/2}
d\tau d\tau' q(\tau) K(\tau - \tau') q(\tau')
\label{act_quantro}
\end{eqnarray}
where $h_{s_1,s_2}$ is the step function being unity in the
interval $[s_1,s_2]$ and zero anywhere else and the kernel
follows from (\ref{1overfaction}). For the two flip
contribution the action also depends on  the flip times
$s_1$ and $s_2$. The influence kernel $K(\tau)$ gives
according to the analysis of Sec.~\ref{kernel} rise to a
static and a dynamical contribution, see
(\ref{1overfaction}). We first concentrate on weak friction
for both,  such that a perturbation theory to first order
in $\tilde{J}_s$ and $\tilde{\kappa}$ applies. By resorting
again to a semiclassical evaluation this amounts to the
fact to insert the undamped bounces $q_{\rm b}$ and $q_{\rm
flip}$, respectively, into the influence functional. The
rate enhancement due to Zener flips can then be estimated
by the difference between the simple bounce and the flip
bounce actions, i.e.,
\begin{eqnarray}
\Delta S(\tilde{J}_s,\tilde{\kappa})&=&S_{\rm
b}(\tilde{J}_s=0,\tilde{\kappa}=0)-
S_{\rm flip}(\tilde{J}_s=0,\tilde{\kappa}=0)\nonumber\\
&& \hspace{-2.0cm}-\frac{\tilde{J}_s}{\hbar\beta}\left\{
\left[\int_{-\hbar\beta/2}^{\hbar\beta/2}d\tau q_{\rm
b}(\tau)\right]^2-\left[\int_{-\hbar\beta/2}^{\hbar\beta/2}d\tau
q_{\rm flip}(\tau)\right]^2\right\}\\ \nonumber
&&  \hspace{-2.5cm}+\frac{1}{2}
\left\{\int_{-\hbar\beta/2}^{\hbar\beta/2} d\tau d\tau'
\tilde{k}(\tau - \tau')\left[q_{\rm b}(\tau)  q_{\rm
b}(\tau')- q_{\rm flip}(\tau)  q_{\rm
flip}(\tau')\right]\right\}\, . \label{diffact}
\end{eqnarray}
The explicit form for the flip bounce and its action has
been given in \cite{ankerhold} to which we refer for
further details. Since the flip bounce always exhibits are
smaller amplitude and width (see fig.~\ref{figure-bounce}),
the terms in curly brackets are always positive. Hence, we
arrive at the important result that weak static noise
always lowers the rate enhancement due to Zener flip
tunneling, while weak dynamical noise always increases its
effect. The impact of which actually prevails depends only
on the coupling strengths since the respective tunneling
orbits enter both corrections in the same way. We note that
so far (\ref{diffact}) relies only on the weak coupling
limit, but applies to any kind of spectral density of the
environment.
%------------------------- figure: bounce2  ---------------------

%--------------------- figure: zenerenhance -------------------
\begin{figure}
%\vskip-1.5cm
\includegraphics[width=85mm]{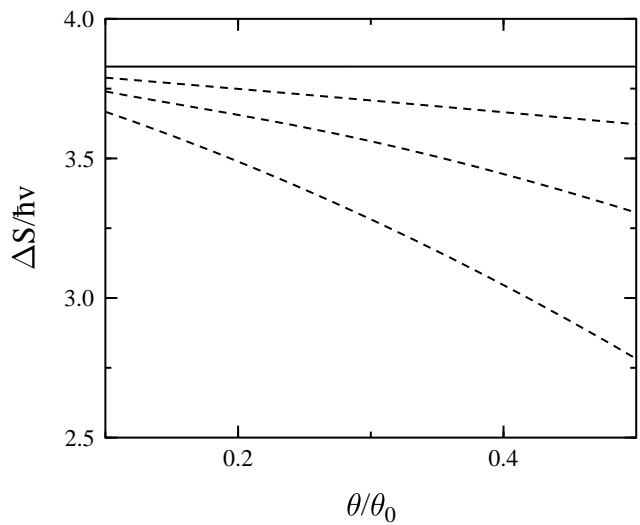}
%\vskip-5.0cm
\caption{Rate enhancement due to Zener flip tunneling vs.
temperature in presence of a bath with a $1/f$ spectrum
and various values for the coupling constant (from top to
bottom) $\tilde{j}=0.0\ \mbox{(solid)}, 0.1, 0.2, 0.3$
(dashed).} \label{figure-action}
\end{figure}

As we have seen above, for a $1/f$ noise the impact of
dynamical modes is much smaller than that of the sluggish
ones so that we expect a shrinking of the enhancement. In
order to see that explicitly, the tunneling orbits in
presence of damping are determined numerically.
Particularly,  the flip-bounce is determined by a
vanishing variation of $S[q;s_1,s_2]$ with respect to all
periodic paths and their corresponding flip times. It
turns out that  the optimal spin flips occur just as the
trajectory reaches the intersection point and that they
are centered symmetrically around $\tau=0$. For the
stationary action the time dependent potential can
therefore be replaced by a piecewise potential $V_c
(q)=V_-(q) \Theta(q_c-q)+V_+(q) \Theta(q-q_c)$ where $q_c$
denotes the intersection point. The equation of motion for
the flip bounce thus reads
\begin{equation}
-M\ddot{q}_{\rm flip}(\tau)+V_c'(q_{\rm
flip})=\int_{-\hbar\beta/2}^{\hbar\beta/2} d\tau'\,
K(\tau-\tau')\, q_{\rm flip}(\tau')\ .
\end{equation}

Unfortunately, the straightforward approach developed for
standard MQT rates and exploited also in
Sec.~\ref{adiabatic}, where the equation of motion is
solved numerically  by switching to Fourier space, does
not work here due to the piecewise potential. Its
derivative gives rise to $\delta$ function contributions,
which pose numerical instabilities that cannot be
circumvented by the usual rescaling. We thus determined
$q_{\rm flip}$ in coordinate space  in an iterative
procedure. One starts with the known flip-bounce for
vanishing dissipation $q_{\rm flip}^{(0)}$ on the right
hand side and calculates the dissipative flip-bounce
$q_{\rm flip}^{(k)}$ by successively using $q_{\rm
flip}^{(k-1)}$ for the dissipative part. The final
flip-bounce follows after a sufficient number of
iterations and by preserving the boundary conditions
$q_{\rm flip}(\tau\to \infty)=0$ for $\hbar\beta\to
\infty$. Typically, convergence is achieved quickly after
$5-10$ steps depending on the dissipation strength. From
the flip-bounce the corresponding action is obtained
which, for weak friction, leads together with the
non-dissipative prefactor $f_{\rm flip}(\Delta_c)$ and the
standard MQT rate to the total rate in the flip range.

\begin{figure}
%\vskip-4cm
\includegraphics[width=85mm]{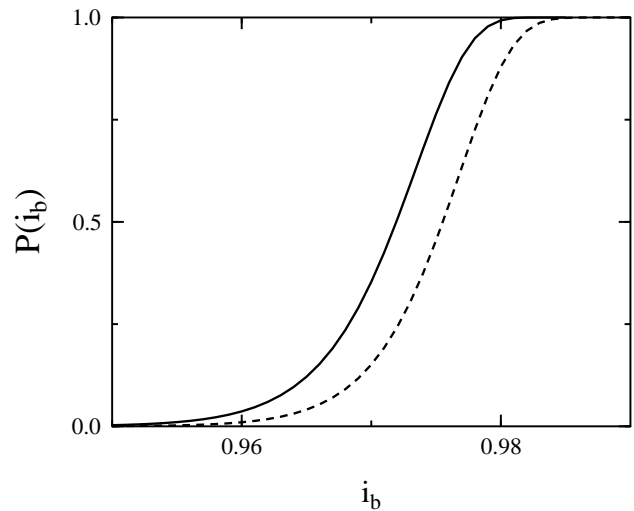}
%\vskip-6cm
\caption{Probability to escape vs.\ external
bias current for the quantronium circuit and in the
parameter range where Zener flip tunneling occurs. The
solid line is the result including Zener flips, while the
dashed one is generated using the standard MQT rate.}
\label{figure-zeners}
\end{figure}

In fig.~\ref{figure-action} the rate enhancement due to
Zener flip tunneling given by the action difference
between standard bounce action and flip bounce action is
shown for various values of the coupling to a $1/f$ bath.
It is clearly seen that the impact of the static modes
prevails and suppresses the enhancement with increasing
temperature according to the above discussion. For weak
coupling, however, the suppression remains small even at
higher temperatures. To further illustrate the Zener flip
effect on rate measurements, we show in
fig.~\ref{figure-zeners} the probability to escape
$P(i_b)$ with varying bias current $i_b$ [see
(\ref{probi})] for the quantronium circuit and in the
range where Zener flips occur. Qualitatively, Zener flips
lead to a pronounced shift of the s-curve towards smaller
values of the external bias current, though, with a smaller
slope in the first part of the curve.  The influence of
sluggish bath modes is to slightly move the Zener-curve
back towards the standard MQT curve, but for weak damping
this effect is hardly visible in the s-curves. By
comparing with fig.~\ref{slow-s-curves}, however, one
realizes that the Zener-curve resembles a standard MQT
curve in presence of a stronger low frequency environment.
Experimentally, some further information about the
influence of environmental modes is thus required to
detect the Zener flip effect in a single s-curve
measurement. The effect is clearly observable when one
measures the bias current needed to maintain a certain
value for $P(i_b)$ while sweeping through the Zener flip
range by varying the magnetic flux $\phi$ \cite{zener2}.
%-----------------------figure: zenerscurves ------------------

\section{Conclusions}
Macroscopic quantum tunneling has regained new interest in
the context of quantum information processing based on
Josephson junction circuits. In this paper we analyzed the
influence of low to moderate frequency noise on the
tunneling rate out of a metastable well, where the noise
spectrum is restricted to frequencies somewhat smaller
than the typical frequency for oscillations around the
well bottom (plasma frequency). A sluggish bath leads to a
rate enhancement linear in the temperature, in contrast to
the typical $T^2$ behavior for ohmic environments. For a
sluggish bath with an additional $1/f$ characteristic it
turns out that even in
 presence of dynamical modes the static component by far
 prevails. This verifies that for MQT processes $1/f$ noise can basically
be treated as classical noise and in an adiabatic
approximation where for very weak coupling one effectively
averages over rates for an ensemble of potential barriers.

Based on these findings the impact of noise on a particular kind
of MQT process has been studied, where a particle carrying
a spin-$1/2$ as an  internal degree of freedom experiences
a Zener transition while tunneling.  By comparing this new
decay channel with ordinary MQT, we find that weak static
noise suppresses the enhancement of Zener flip tunneling,
while weak dynamical noise increases it. Further, we find
that for escape probabilities (s-curves) the Zener flip
effect leads qualitatively to a similar result as a
standard MQT process under the stronger influence of
sluggish bath modes. For the experimental observation of
Zener flip tunneling it is thus advantageous to have an
environment weakly coupled to the system (higher
$Q$-factor) and a spectrum with a decreasing intensity
towards low frequency modes, e.g.\ with a prevailing ohmic
characteristic.

\section*{Acknowledgements}
This work benefitted from fruitful discussions with E.
Collin, D. Esteve, D. Vion, and H. Grabert. JA
acknowledges a Heisenberg fellowship of the DFG.
\vskip0.4cm
$^\dagger$ Present address: Department of Physics and Astronomy, University of Basel, Klingelbergstrasse 82, CH-4056 Basel, Switzerland

\end{document}